\documentclass[manuscript]{aastex}

\begin{document}


\title{Neutral interstellar helium parameters based on Ulysses/GAS and IBEX-Lo observations: \\
       what are the reasons for the differences? }


\author{O. A. Katushkina\altaffilmark{1} and V. V. Izmodenov\altaffilmark{1,2,3}}
\affil{Space Research Institute of Russian Academy of Sciences, Moscow, Russia}
\email{okat@iki.rssi.ru}

\author{B. E. Wood\altaffilmark{4}}
\affil{Naval Research Laboratory, Space Science Division, Washington DC, USA}

\and

\author{D. R. McMullin\altaffilmark{5}}
\affil{Space Systems Research Corporation, Alexandria, USA}


\altaffiltext{1}{Space Research Institute of Russian Academy of Sciences, Moscow, Russia.}
\altaffiltext{2}{Lomonosov Moscow State University, Moscow, Russia.}
\altaffiltext{3}{Institute for Problems in Mechanics of Russian Academy of Sciences, Moscow, Russia.}
\altaffiltext{4}{Naval Research Laboratory, Space Science Division, Washington DC, USA}
\altaffiltext{5}{Space Systems Research Corporation, Alexandria, USA}


\begin{abstract}
Recent analysis of the interstellar helium fluxes measured in 2009-2010 at Earth orbit by the Interstellar Boundary Explorer (IBEX) has suggested that the interstellar velocity (both direction and magnitude) is inconsistent with that derived previously from Ulysses/GAS observations made in the period from 1990 to 2002 at 1.5-5.5 AU from the Sun. Both results are model-dependent and models that were used in the analyses are different. In this paper, we perform an analysis of the Uysses/GAS and IBEX-Lo data using our state-of-the-art 3D time-dependent kinetic model of interstellar atoms in the heliosphere. For the first time we analyze Ulysses/GAS data from year 2007, the closest available Ulysses/GAS observations in time to the IBEX observations.

 We show that the interstellar velocity derived from the Ulysses 2007 data is consistent with previous Ulysses results and does not agree with the velocity derived from IBEX.
This conclusion is very robust since, as is shown in the paper, it does not depend on the ionization rates adopted in theoretical models.

 We conclude that Ulysses data are not consistent with the new LISM velocity vector from IBEX.
In contrast, IBEX data, in principle, could be explained with the LISM velocity vector derived from the Ulysses data. This is possible for the models with the interstellar temperature increased from 6300 K to 9000 K. There is a need to perform further study of possible reasons for the broadening of the helium signal core measured by IBEX. This could be an instrumental effect or due to unconsidered physical processes.
\end{abstract}


\section{Introduction}

 The Solar System is surrounded by the partially ionized plasma of the Local Interstellar Medium (LISM).
 The most abundant neutral component in the LISM is atomic hydrogen.
  Minor neutral components in the LISM are atomic helium, oxygen, nitrogen, and others.
 The Sun is moving through the LISM with a relative velocity about 20-30~km/s. The supersonic solar wind (SW) interacts with the charged component
 of the interstellar plasma and the result is the SW/LISM interaction region, which is called the heliospheric interface \citep{bm_1993}.
 The mean free path of interstellar neutrals is comparable to the size of the heliospheric interface \citep[see e.g.][]{izmod_etal_2001}. Therefore, neutral atoms penetrate
 through this region into the heliosphere, where they can be measured directly or indirectly.

 Being measured in the heliosphere, the interstellar neutrals are the main source of information on the LISM parameters, because charged LISM particles are deflected by the solar wind and do not enter the heliosphere. Although hydrogen (H) atoms have the largest number density of interstellar neutrals, they are not the easiest to study from inside the heliosphere, because during their motion through the heliospheric interface H atoms interact with the interstellar and solar wind protons by charge exchange
  ($H+H^{+} \leftrightarrow H^{+}+H$). As a result, new so-called secondary interstellar H atoms are created, and their distribution function depends on local plasma parameters. Therefore, hydrogen distributions in the heliosphere (e.g. at the heliospheric termination shock) are considerably disturbed compared with the original distribution in the LISM \citep[see][]{izmod_etal_2001}. In addition, near the Sun hydrogen atoms are affected by substantial solar radiation pressure, varies with time and the velocity of particles, which results in more complications for modelling of hydrogen distribution
  as compared with helium (radiation pressure is negligible for helium).
 That is why it becomes challenging to use the hydrogen distribution, for example at 1 AU, to derive the LISM parameters, because one needs to
 take into account perturbation of the hydrogen parameters in the heliospheric interface \citep{kat_izmod_2010, kat_izmod_2011}.

On the other hand,
 due to the charge exchange interactions, interstellar hydrogen distributions inside the heliosphere can be used as remote diagnostics of the heliospheric interface. Since 1970s, interstellar hydrogen in the heliosphere has been studied remotely by numerous measurements of backscattered solar Lyman-alpha radiation
  by, e.g., OGO-5 \citep{thomas_krassa_1971, bertaux_blamont_1971}, Prognoz-5 and 6 \citep{bertaux_etal_1985}, SOHO/SWAN \citep{costa_etal_1999, quemerais_izmod_2002}, Voyager-1/2 \citep{quemerais_etal_2010}, Hubble Space Telescope \citep{vincent_etal_2011} and others. Nowadays interstellar hydrogen atoms for the first time are measured directly at Earth orbit by the IBEX-Lo sensor on board the Interstellar Boundary Explorer (IBEX) spacecraft. Some data and results of these observations are presented in \citet{saul_etal_2012} and \citet{schwadron_etal_2013}.

  Contrary to the H atoms, it is known that interstellar helium (He) atoms penetrate into the heliosphere almost freely. They only weakly interact with protons ($H^{+}$) and helium ions ($He^{+}$) by charge exchange, due to small charge exchange cross sections
  \citep[see, e.g., section~7 in][]{bzowski_etal_2012}.
  This means that measurements of the interstellar helium near the Sun
 can be used to determine the temperature ($T_{LISM}$) and relative velocity vector ($\textbf{V}_{LISM}$) of the LISM.
 Inside the heliosphere, the interstellar helium flow
  suffers from effects of solar photoionization and electron impact ionization. Rates of these processes are partially known from
  different observations of the solar irradiance and the solar wind \citep{mcmullin_etal_2004, bzowski_etal_2012}. So, to obtain
  the LISM parameters from the local observations inside the heliosphere one should use
 a theoretical model of interstellar helium distributions in the heliosphere, which takes into account all important ionization processes, and then
 solves the inverse problem to find the LISM parameters providing the best agreement between results of the numerical modeling and the experimental data.

  Such a technique to derive the LISM parameters from the interstellar helium measurements in the heliosphere was applied to data
  from the GAS instrument on board the Ulysses spacecraft \citep{banaszkiewicz_etal_1996, witte_etal_1993, witte_etal_1996, witte_2004}. The Ulysses/GAS instrument was designed for direct measurements of interstellar helium.  These measurements were performed from 1990 to 2007. Analysis of the Ulysses/GAS data from 1990 to 2002 by \citet{witte_2004} yielded
  the following LISM parameters: number density of interstellar helium
  $n_{He,LISM}=0.015 \pm 0.003$~cm$^{-3}$, temperature $T_{LISM}=6300 \pm 340$~K, relative SW/LISM velocity $V_{LISM}=26.3 \pm 0.4$~km/s, and
  direction of the interstellar wind in J2000 ecliptic coordinates at longitude $\lambda_{LISM}=75.4^{\circ} \pm 0.5^{\circ}$ and latitude
  $\beta_{LISM}=-5.2^{\circ} \pm 0.2^{\circ}$. These parameters were found to be consistent with other experimental data \citep{moebius_etal_2004,
  lallement_etal_2004, vallerga_etal_2004} and remained canonical until recently.

  In October 2008 a new NASA mission, IBEX, was launched \citep{mccomas_etal_2009}.
  The main goal of IBEX is to study the three-dimensional structure of the heliosphere using measurements
of heliospheric neutrals (hydrogen, helium and oxygen) in different energy channels \citep{mccomas_etal_2009, moebius_etal_2009}.
IBEX is primarily designed to study high energy neutrals formed by charge
  exchange between the termination shock and the heliopause rather than LISM neutrals, but the IBEX-Lo sensor is also
  capable of observing the LISM neutrals at certain times of the year.
IBEX-Lo \citep{fuselier_etal_2009} is designed to measure the low-energy neutrals in the energy range from 0.01 to 2 keV.

IBEX-Lo measurements of interstellar helium in 2009-2010 were analyzed recently by \citet{bzowski_etal_2012} and \citet{moebius_etal_2012}.
The analysis of \citet{bzowski_etal_2012} was based on a model of the helium distribution similar to that of \citet{witte_2004}, but
 taking into account more recent data on the helium ionization rates. \citet{moebius_etal_2012} have performed an analytical analysis of the
  IBEX-Lo measurements in the context of a stationary and axisymmetric model \citep[the so-called ``classical hot model'', see,][]{meier_1977, wu_judge_1979, lallement_etal_1985, lee_etal_2012}.The following LISM parameters were obtained as the result of these investigations: $T_{LISM}=6300$~K, $V_{LISM}=23.2$~km/s, $\lambda_{LISM}=79^{\circ}$, $\beta_{LISM}=-4.98^{\circ}$. These mean values were taken from \citet{mccomas_etal_2012}, who used weighted means to combine the two independent results of \citet{bzowski_etal_2012} and \citet{moebius_etal_2012}. The IBEX-Lo analysis of possible values of the interstellar parameters ($V_{LISM}, \lambda_{LISM}, \beta_{LISM}, T_{LISM}$) suggests a ``tube'' of allowable fits in the 4D parameter space. This ``tube'' is characterized by 1) uncertainties that represent the widths of the ``tube'', and 2) bounding ranges that characterize the length of the ``tube''
\citep[see][for details]{mccomas_etal_2012}. The uncertainties and the bounding ranges are shown in Table~1 of \citet{mccomas_etal_2012}.


 The velocity of the interstellar flow obtained from the IBEX-Lo data
is about 3~km/s less and its direction 4$^{\circ}$ different compared with the previous results of \citet{witte_2004}.
Note that results of both \citet{witte_2004} and \citet{mccomas_etal_2012} are model-dependent, and different models have been used. Therefore, it is worthwhile to analyze both GAS and IBEX-Lo data in the context of one model.

Although the differences in the LISM parameters may not seem large, they may actually be physically significant.
For example, the low velocity measurement from IBEX has stimulated a debate about the existence of the Bow Shock \citep{mccomas_etal_2012, zank_etal_2013}.
Also, changes in the $\textbf{V}_{LISM}$ direction influence the orientation of the hydrogen deflection plane (HDP) \citep{lallement_etal_2005, lallement_etal_2010},
which in turn leads to a different inferred configuration of the interstellar magnetic field within the HDP. Changes in the SW/LISM relative velocity could also affect the position of the heliopause (i.e. the contact discontinuity where dynamic pressure of the
interstellar plasma and the solar wind are equal to each other). This is very important for interpreting data from the Voyager spacecraft, which are approaching the
heliopause. Voyager~1 in fact may have already crossed the heliopause \citep{gurnett_etal_2013}.

In this paper, we perform an analysis of both Uysses/GAS data (in years 2001 and 2007) and IBEX-Lo data (in year 2009) using our state-of-the-art 3D time-dependent kinetic model of interstellar atoms in the heliosphere. We do not aim to repeat the detailed analyses performed previously by Witte and Bzowski et al., and restrict ourselves to a few individual observations from both spacecraft. We provide the first analysis of Ulysses/GAS data obtained in 2007, which is closer in time to the observations of IBEX. Calculations were performed for the ``old'' and ``new'' LISM velocity vectors, and for different ionization rates adopted in the model.
We explore the role of the ionization rates on the differences in the LISM velocity vector obtained from the GAS and IBEX-Lo data.

\section{Model of the interstellar helium distribution in the heliosphere}

The kinetic equation is solved to obtain the distribution of the interstellar helium atoms in the heliosphere:
\begin{eqnarray}
   \frac{\partial f(\textbf{r},\textbf{w},t)}{\partial t}
   + \textbf{w}\cdot\frac{\partial f (\textbf{r},\textbf{w},t)}{\partial \textbf{r}}+ & \nonumber  \\
   \frac{\textbf{F}_{g}(r)}{m_{He}}\cdot \frac{\partial f(\textbf{r},\textbf{w},t)}{\partial \textbf{w}} &
   = -\beta_{ph}(r,t,\lambda)\cdot f(\textbf{r},\textbf{w},t).\label{Boltzmann}
\end{eqnarray}
Here, $f(\textbf{r},\textbf{w},t)$ is the velocity distribution function, $\textbf{w}$ is the velocity vector of a He atom, $\beta_{ph}(r,t,\lambda)$ is the
 photoionization rate, and $\textbf{F}_{g}$ is the gravitational force.  The gravitational force can be written in the form
\begin{equation}
\textbf{F}_{g}(\textbf{r})=- \frac{G m_{He} M_{s}}{r^{2}}\cdot \frac{\textbf{r}}{r},\label{force}
\end{equation}
where $G$ is the gravitational constant, $m_{He}$ is the mass of He atom, and $M_s$ is the mass of the Sun.  We assume that $\beta_{ph}(r,t,\lambda)\sim 1/r^2$, i.e.
\begin{equation}
  \beta_{ph}(r,t,\lambda)=\beta_{ph,E}(t,\lambda)\cdot \frac{r_E^2}{r^2} \label{force}
\end{equation}
where $r_E=1$~AU and $\beta_{E}(t,\lambda)$ is the photoionization rate at 1 AU, which in general depends on time and
heliolatitude ($\lambda$) due to spatial and temporal variations of the solar EUV irradiance.

The outer boundary of our computational region is a Sun-centered sphere with radius 1000~AU. At this distance from the Sun the influence of
solar gravitation and photoionization is negligibly small. Therefore, we assume that the interstellar helium flow is undisturbed, and
its velocity distribution function is a simple Maxwellian with pristine LISM parameters:
\begin{equation}
  f_{M}(\textbf{w})=  \frac{n_{He, LISM}}{\pi \sqrt{\pi} \cdot c_{s}^{3} }
 \cdot \exp \left( -\frac{\left(\textbf{V}_{LISM}-\textbf{w} \right)^{2}}{c_{s}^{2}} \right)  \mbox{,} \quad
c_{s}=\sqrt { \frac{2k_{b}T_{LISM}}{m_{He}} }, \label{maxwell}
\end{equation}
where $n_{He, LISM}$ is number density of neutral helium in the LISM, $T_{LISM}$ is temperature of the LISM,
$\textbf{V}_{LISM}$ is the velocity vector of the LISM relative to the Sun, and $k_b$ is Boltzmann's constant.
The direction of $\textbf{V}_{LISM}$ is defined by two spherical angles (latitude $\lambda_{LISM}$ and longitude $\beta_{LISM}$) in the solar
ecliptic (J2000) coordinate system.
In our calculations, parameters of the LISM were taken either from the new results of \citet{mccomas_etal_2012} based on IBEX-Lo data, or
from previous results of \citet{witte_2004} based on Ulysses data.

Equation~(\ref{Boltzmann}) does not take into account the solar radiation force caused by scattering of solar photons on atoms, which
is negligible for helium. We also neglect electron impact ionization and charge exchange ionization, because the dominant loss process
is photoionization by solar EUV radiation.  For example,
at 1~AU in the ecliptic plane the electron impact ionization rate is approximately equal to 1-2$\cdot 10^{-8}$~s$^{-1}$
\citep{mcmullin_etal_2004, bzowski_etal_2012}, while the typical value of the photoionization rate at 1~AU is about $1.5 \cdot 10^{-7}$~s$^{-1}$ at solar maximum and
$5.5 \cdot 10^{-8}$~s$^{-1}$ at solar minimum \citep{bzowski_etal_2012}. The photoionization rate decreases with heliocentric distance
as $~1/r^2$, while the electron impact ionization rate decreases with distance from the Sun much faster than $1/r^2$ due to rapid cooling of the solar wind electrons \citep{mcmullin_etal_2004, bzowski_etal_2012}.
Thus, electron impact ionization may be important only very near the Sun at $<$~1~AU.
Another loss process for helium atoms that we ignore is charge exchange with the solar wind protons and solar wind alpha-particles, as
the charge exchange rate at 1~AU is only about 4~\% of the typical photoionization rate \citep{bzowski_etal_2012}.

Equation~(\ref{Boltzmann}) is a linear partial-differential equation and can be solved by a method of characteristics either in an axisymmetric (2D) stationary case
(with constant photoionization rate), or in a three-dimensional (3D) time-dependent case, where the photoionization rate depends on heliolatitude and time.
In our calculations we experimented with both 2D stationary and 3D time-dependent models. The 3D time-dependent ionization rate is described in Appendix~B.

\section{Modeling of the interstellar helium fluxes measured by Ulysses/GAS in 2001 and 2007}

\begin{figure}
\plotone{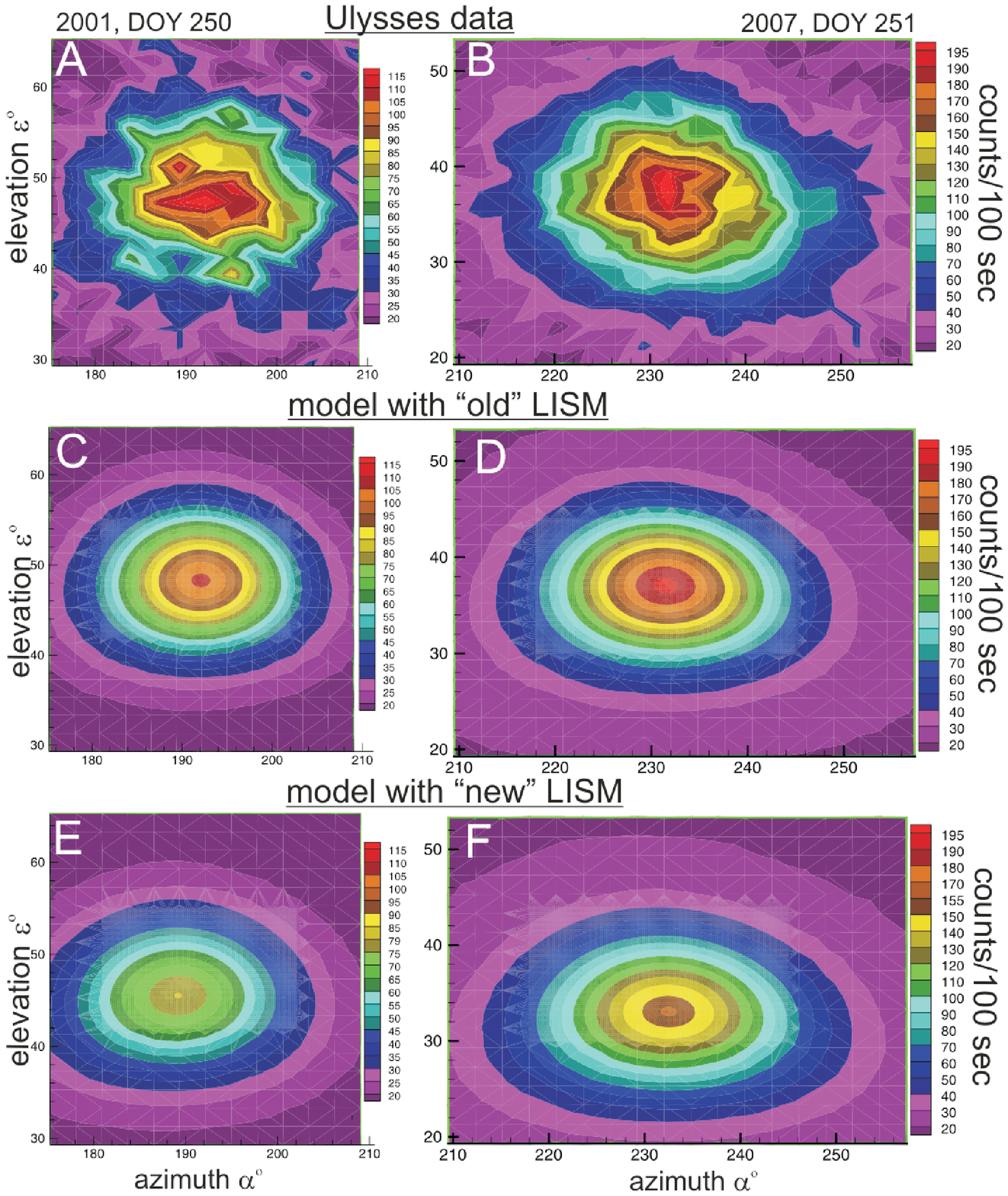}
\caption{A-B: Sub-images extracted from the GAS data.
C-D: results of the model calculations with the ``old'' LISM velocity vector.
E-F: results of the model calculations with the ``new'' LISM velocity vector.
These results were obtained using an axisymmetric stationary model with the constant ionization rate ($\beta_{ph,E}=1.5 \cdot 10^{-7}$~s$^{-1}$
in 2001, and $\beta_{ph,E}=7 \cdot 10^{-8}$~s$^{-1}$) in 2007.} \label{maps_2Dstat}
\end{figure}

In this section, results for Ulysses/GAS data are presented. Technical
details of the flux calculations are described in Appendix~A.
We considered two maps of the interstellar helium fluxes obtained using the Narrow Field of View (NFOV) detector of Ulysses/GAS, one in 2001 (day of year or DOY is 250) and one in 2007 (DOY is 251).
The angular resolution of the GAS measurements for these moments of time is $\Delta \alpha=2.8^{\circ}$ and $\Delta \epsilon = 2^{\circ}$ (azimuth $\alpha$ and elevation $\epsilon$ are angles defined by the direction of the line of sight in the spacecraft's system of coordinates).
The parts of sky maps containing the He beam are presented in Fig.~\ref{maps_2Dstat}~A and B.

We calculate synthetic maps using our model of the helium distribution, with two sets of the LISM parameters.
The first set is based on the analysis of Ulysses/GAS data by \citet{witte_2004}, which we will refer to as the ``old'' LISM parameters.
The second set is based on the recent analysis of IBEX-Lo data by \citet{mccomas_etal_2012}, \citet{bzowski_etal_2012} and \citet{moebius_etal_2012}, which
we will refer to as the ``new'' LISM parameters.

We first compute results using the simplified axisymmetric stationary model.
In this model we assume that the photoionization rate at 1~AU ($\beta_{ph,E}$) is constant and equal to the following values \citep{bzowski_etal_2012}:
 $\beta_{ph,E}=1.5 \cdot 10^{-7}$~s$^{-1}$ in 2001 and
 $\beta_{ph,E}=7 \cdot 10^{-8}$~s$^{-1}$ in 2007.

Fig.~\ref{maps_2Dstat} presents the results, in which the observed background in the GAS data was artificially added to the
 model results (see Appendix~A).
 Quantitative differences in the absolute values of counts between the models and data can be explained by assumptions made
about the photoionization rate and the helium number density in the LISM, but we are not interested in the absolute values of counts here.

We focus on the direction (or position) in the sky where the helium flux is at maximum (i.e. the center of the He beam),
because this direction is most sensitive to  $\textbf{V}_{LISM}$, and therefore appropriate to distinguish between
 the ``new'' and ``old'' LISM velocity vectors.
The directions of the center of the He beam are presented in Table 1 for the GAS data ($\textbf{r}_{0,data}$) and for the model ($\textbf{r}_{0,model}$).
Each direction is characterized by two angles: $\alpha_0$ and $\epsilon_0$.
It is important to note that the direction of the center of the He beam
can be determined from the GAS data only with the precision of $\pm 1.4^{\circ}$ for $\alpha$ and $\pm 1^{\circ}$ for $\epsilon$
due to the limited angular resolution (see Appendix A).
In the model, the direction of the He beam center depends on the computational grid resolution and can be determined with higher precision compared to the data.
In our calculations, the precision is $\pm 0.1^{\circ}$ for both $\alpha$ and $\epsilon$.

\begin{figure}
\plotone{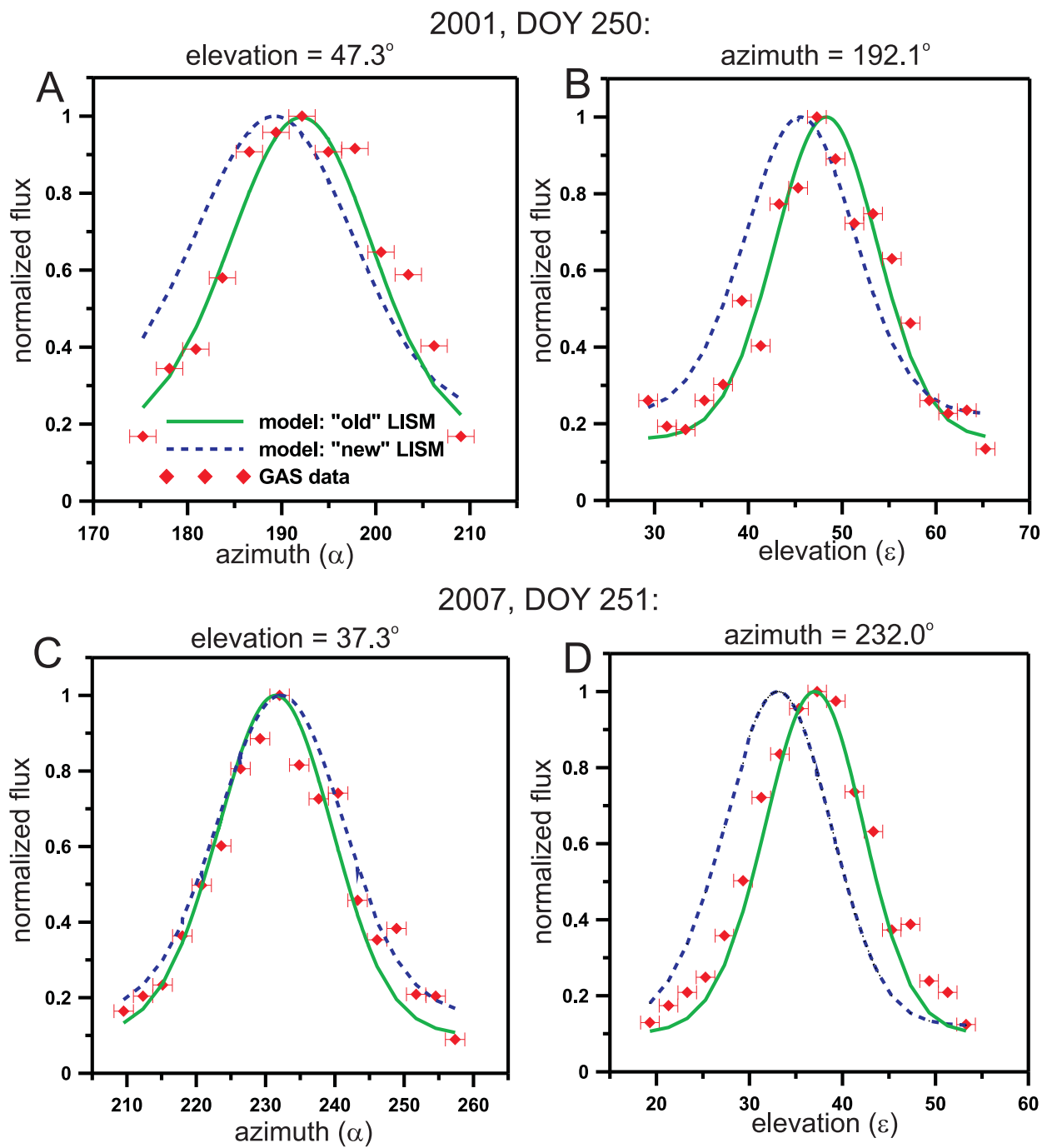}
\caption{Normalized helium fluxes are shown as a function of azimuth for the beam-center elevation (A, C), and as a function of
elevation for the beam-center azimuth (B, D). Plots A and B correspond to the 2001 map from Fig.~1, and plots C and D correspond to 2007. Red symbols represent the GAS data,
green solid curves the model results with the ``old'' LISM velocity vector, and blue dashed curves the model results with the ``new'' LISM velocity vector.} \label{GAS_1Dstat}
\end{figure}

To illustrate the results further, we present in Fig.~\ref{GAS_1Dstat} 1D plots of the normalized fluxes through the elevation and azimuth angles that define the beam center for the GAS data.
For this figure the fluxes are normalized to have a maximum of 1. Error bars ($\pm 1.4^{\circ}$ for azimuth and $\pm 1^{\circ}$ for elevation) are added to the points of GAS data in the plot. It is clearly seen from Fig.~\ref{GAS_1Dstat} that models with different
LISM velocity vectors give different beam locations.

Table 1 additionally shows values of $\gamma$, which measures the deviation between the model
and data helium beam directions: $\gamma = \arccos (\textbf{r}_{0,data}\cdot \textbf{r}_{0, model})$.
For the model with the ``new'' LISM velocity, $\gamma = 2.73$ in 2001 and $\gamma = 4.29$ in 2007, discrepancies that are far larger
than the precision of the direction determined from the data. We conclude that the He beam directions obtained for the model assuming the ``new'' LISM velocity
deviates from GAS observations by several degrees.
In contrast, the directions of the He beam obtained by the model with the ``old'' LISM velocity agree with the GAS data much better.
Values of $\gamma$ are 1.1 and 0.47 for 2001 and 2007, respectively, which are less than the angular resolution of Ulysses/GAS data.

The above results were obtained using the axisymmetric stationary model under the assumption of a photoionization rate varying as 1/$r^2$ and independent of time and heliolatitude.
However, space measurements show that the photoionization rate does vary with time and heliolatitude.
Appendix~B describes the temporal and heliolatitudinal variations of the photoionization rate based on several spacecraft data.
To explore the role of these variations on the helium flux maps we performed the model calculations using our 3D code and a time and heliolatitude dependent photoionization rate.
The maps obtained in this calculations are not shown here, because they are qualitatively similar to those shown in Figure 1.
The main difference with the presented maps is in the absolute fluxes, which are beyond the scope of this paper.
The angle $\gamma$ is 1.09  and 0.47 for 2001 and 2007, respectively, for the model with the ``old'' LISM velocity.
For the model with the ``new'' LISM velocity: $\gamma=2.73$ in 2001 and $\gamma=4.29$ in 2007.
These values are almost the same as in the axisymmetric stationary case.
We conclude that variations of the photoionization rate are not important for the direction of the He beam.
It is also possible to show \citep{katushkina_etal_2014} that this direction does not depend on the LISM temperature.

Thus, our conclusions from the simple axisymmetric stationary model are the same as in the more general 3D and time-dependent case.
This means that for any assumed ionization rate and LISM temperature,
the model with the ``new'' LISM velocity vector cannot reproduce the position of the He beam in Ulysses/GAS data either in 2001 or in 2007, while the model with the ``old'' LISM velocity nicely reproduces the GAS data for both considered maps.
Our analysis also shows that Ulysses data obtained in 2007 are not suggestive of a change in the LISM velocity vector from the 1996-2004 Ulysses/GAS data.

\section{Modeling of the interstellar helium fluxes measured by IBEX-Lo}

A detailed description of the IBEX-Lo sensor can be found in \citet{fuselier_etal_2009}, and \citet{moebius_etal_2009, moebius_etal_2012}.
Technical details of the modelling of the interstellar helium fluxes measured by IBEX-Lo are presented in Appendix~C. Here we will give only a brief
description of the geometry of IBEX observations.

\begin{figure}
\plotone{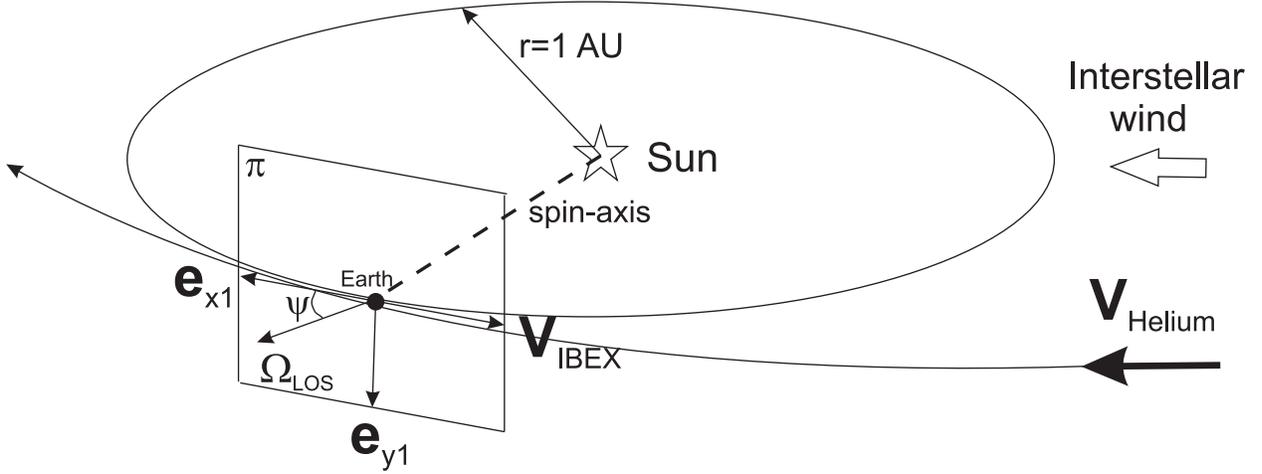}
\caption{ Geometry of IBEX observations. The dashed line shows the spin-axis of IBEX, which is pointed approximately toward the Sun (in the calculations, we use
real directions of the spin-axis from the ISOC database). Plane $\pi$ is perpendicular to the spin-axis, and IBEX performs measurements in this plane
(i.e. all line of sight $\Omega_{LOS}$ belong to plane $\pi$). Plane $\pi$ is formed by two orthogonally related vectors: $\textbf{e}_{x1}$ and $\textbf{e}_{y1}$, both
of them perpendicular to the spin-axis. Vector $\textbf{e}_{x1}$ belongs to the solar ecliptic plane and $\textbf{e}_{x1}\cdot \textbf{V}_{IBEX} <0$;
$\textbf{e}_{y1}=\textbf{e}_{x1}\times \textbf{r}_{Earth-Sun}$. Each line of sight $\Omega_{LOS}$ can be characterized by one angle $\psi$ measured
in plane $\pi$ from $\textbf{e}_{x1}$.
}\label{ibex_geometry}
\end{figure}

IBEX is a satellite in a highly elliptic orbit around Earth. Each orbit takes approximately eight days.
IBEX is a spinning spacecraft, with its spin-axis reoriented toward the Sun at the beginning of each orbit, and the direction of the spin-axis kept the same
during the orbit. IBEX measures the fluxes of interstellar neutrals in the plane perpendicular to the spin-axis (plane $\pi$ in Fig.~\ref{ibex_geometry}).
The IBEX-Lo sensor has a collimator with an angular resolution of $\approx 7^{\circ}$ FWHM (see Appendix~C).
IBEX measures interstellar helium fluxes in January and February of each year. In 2009 this period contains orbits numbered from 13-19, and
we here calculate the interstellar helium fluxes for these orbits.


Typical IBEX-Lo data measure fluxes as a function of angle $\psi$, which is in the plane perpendicular to the spin-axis (see Fig.~\ref{ibex_geometry}).
For each line of sight ($\Omega_{LOS}$), the angle $\psi$ can be replaced by angle $\alpha_{NEP}$, where the NEP-angle
is analogous to the ecliptic latitude, but it is measured from the North Ecliptic Pole.

\begin{figure}
\epsscale{.80}
\plotone{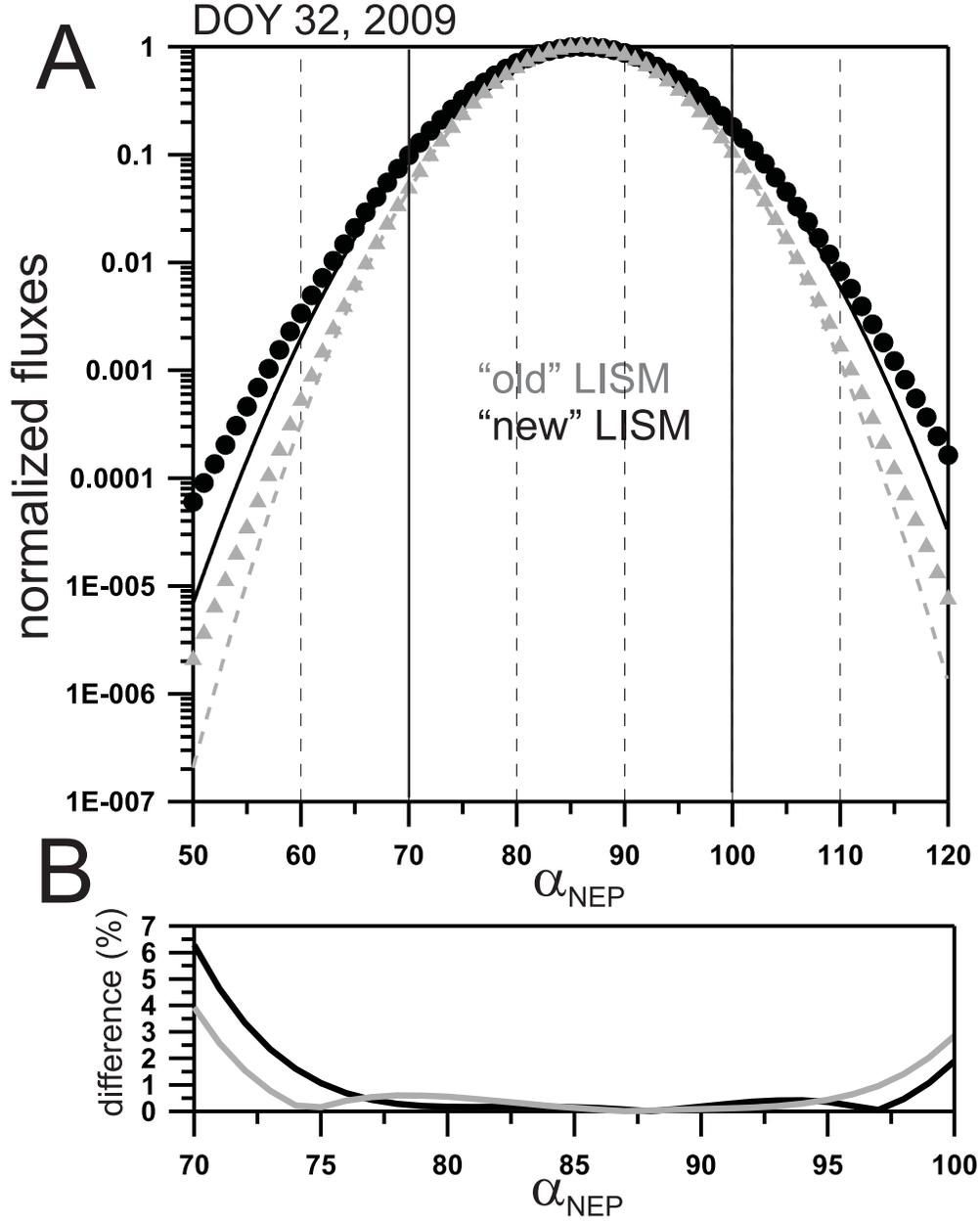}
\caption{A. Normalized interstellar helium fluxes as functions of angle $\alpha_{NEP}$, calculated for DOY 32 in 2009, computed using the stationary axisymmetric model with $\beta_{ph,E}=5\cdot 10^{-8}$~s$^{-1}$, assuming both the ``old'' (grey curve) and ``new'' (black curve)
LISM velocities. Symbols (circles and triangles) show the results of the numerical modelling, while solid and dashed lines show Gaussian functions
fitted to the model results. B. The relative differences between the calculated fluxes and fitted Gaussians for both models.
}\label{ibex_flux_nep}
\end{figure}

We have performed calculations of the helium fluxes for all dates corresponding to IBEX's orbits 13-19, namely for the days of year (DOY) 9-62 in 2009. Calculations were performed using the kinetic model described above with the ``old'' and ``new'' LISM velocity vectors.
Fig.~\ref{ibex_flux_nep} presents an example of the results. This shows normalized fluxes as a function of $\alpha_{NEP}$ at DOY 32 (the first day of orbit 16).
Here we have assumed a constant (with time and heliolatitude) photoionization rate at Earth orbit ($\beta_{ph,E}=5\cdot 10^{-8}$~s$^{-1}$).

We fit the fluxes with a Gaussian core:
\begin{equation}
  f_{G}(\alpha_{NEP} )=f_{max} \exp \left(-\frac{(\alpha_{NEP}-\alpha_{NEP,max})^2}{\sigma^2} \right). \label{f_G}
\end{equation}
This function has three parameters: peak height $f_{max}$, NEP angle of peak $\alpha_{NEP,max}$, and peak width $\sigma$.
Fig.~\ref{ibex_flux_nep}A compares the modeled fluxes (circles and triangles) and Gaussian fitted functions (solid curves). It is seen that the fit is
very good for $\alpha_{NEP}\in[70^{\circ},100^{\circ}]$. Fig.~\ref{ibex_flux_nep}B
 shows the relative difference (in percents) between the calculated fluxes and the Gaussian fits. The discrepancy for $\alpha_{NEP}\in[70^{\circ},100^{\circ}]$ is less than 10~\%. Therefore we use only this interval of angle $\alpha_{NEP}$ to obtain the fit parameters ($f_{max}, \, \alpha_{NEP,max}, \, \sigma$).
These three parameters are used to study how different effects influence the core of helium fluxes, rather than
considering the numerous plots of fluxes as functions of NEP-angle.

\begin{figure}
\epsscale{1.1}
\plotone{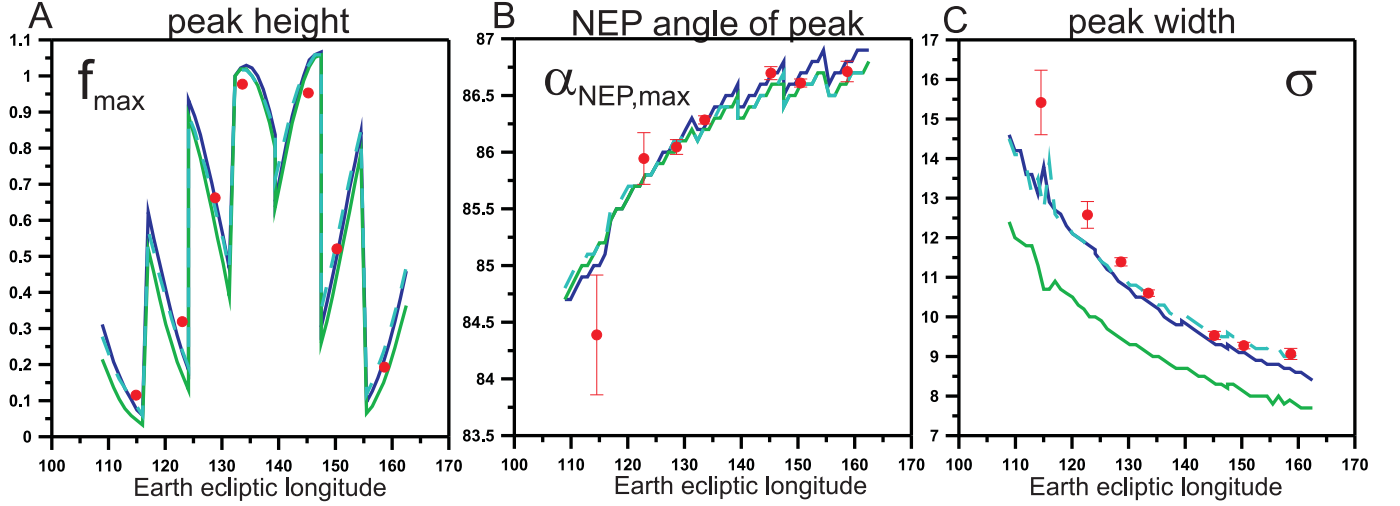}
\caption{The parameters of the Gaussian core: $f_{max}$ is the peak-height (plot A), $\alpha_{NEP,max}$ is NEP-angle of the peak (plot B), $\sigma$ is the peak width (plot C) fitted to the calculated helium fluxes for all days during orbits 13-19 in 2009. All peak heights are normalized to the magnitude of $f_{max}$ at DOY=32.
 Solid green (grey in printed version) curves correspond to the results of modelling with the ``old'' LISM velocity vector, solid blue (black in printed version) curves correspond to the ``new'' LISM velocity vector, and dashed light-blue (light-grey in printed version) curves correspond to the ``old'' LISM velocity vector, but with an enhanced LISM temperature $T_{LISM}=9000$~K.
These results are for the stationary axisymmetrical model with a constant photoionization rate at 1~AU of $\beta_{ph}=5\cdot 10^{-8}$~s$^{-1}$. Red symbols represent the IBEX data (2009) taken from Fig.~9 in \citet{bzowski_etal_2012} (error bars are not shown for peak heights, because they are almost invisible in the figure).
}\label{gauss_ot_DOY}
\end{figure}

The results of modeling and comparison with experimental IBEX data for orbits 13-19 are shown in Fig.~\ref{gauss_ot_DOY}, with IBEX data taken from Fig.~9 of \citep{bzowski_etal_2012}.
The figure presents the parameters of gaussian fits (i.e. ($f_{max}$, $\alpha_{NEP,max}$ and peak width $\sigma$) as a function of Earth ecliptic longitude
(i.e. position of the observer), where
$f_{max}$ is normalized to the $f_{max}$ at DOY=32 (Earth ecliptic longitude $\approx$132$^{\circ}$.) in 2009.
Results are shown for three sets of the
LISM parameters: 1) ``old'' ($V_{LISM}$=26.3~km/s,
$\lambda_{LISM}$=75.4$^{\circ}$, $\beta_{LISM}$=-5.2$^{\circ}$, $T_{LISM}$=6300~K), 2) ``new'' ($V_{LISM}$=23.2~km/s,
$\lambda_{LISM}$=79$^{\circ}$, $\beta_{LISM}$=-4.98$^{\circ}$, $T_{LISM}$=6300~K) and 3) ``old with enhanced temperature'' ($V_{LISM}$=26.3~km/s,
$\lambda_{LISM}$=75.4$^{\circ}$, $\beta_{LISM}$=-5.2$^{\circ}$, $T_{LISM}$=9000~K).


At the first glance, it is seen from Fig.~\ref{gauss_ot_DOY}~A and B that peak height ($f_{max}$) and position of the peak ($\alpha_{NEP,max}$) obtained
from models with  ``old'' and ``new'' LISM parameters are close to each other and to the IBEX data. This is especially important for $\alpha_{NEP,max}$ and means that both the ``old'' and ``new'' LISM velocities give approximately the same result for the direction of maximum helium flux, for the specific geometry of the IBEX measurements.
This is consistent with the previous
analysis performed by \citet{bzowski_etal_2012, moebius_etal_2012} and summarized in \citet{mccomas_etal_2012}. They showed that two sets
of LISM parameters belong to one ``narrow tube'' in 4D space of LISM parameters mentioned above and these two sets lead to
approximately the same position of the peak measured by IBEX.
Recall that in the case of Ulysses, the two models give us different directions of the flux maximum.

If one looks at Fig.~\ref{gauss_ot_DOY}~B carefully it is seen that the IBEX data points (especially for orbits 16 and 17, which correspond to the Earth ecliptic longitude
133$^{\circ}$ and 145$^{\circ}$) appear to align slightly better with the blue curve (model with ``new'' LISM) than the green curve (model with ``old'' LISM).
However, for careful evaluation of distinctions between two models one needs to perform a detailed $\chi^2$-analysis based on least-square method as done
in \citet{bzowski_etal_2012}. In this paper we do not intend to repeat the extensive analysis of \citet{bzowski_etal_2012}, but we focus
on much more pronounced differences between the results of the ``old'' and ``new'' models, which appear in the peak width.

As seen from Fig.~\ref{gauss_ot_DOY}~C
the results of the models with ``old''
and ``new'' sets of LISM parameters are substantially different in the peak width. The peak width is systematically less in the ``old'' LISM velocity model than in the ``new'' model, which agrees well with the IBEX data.
Similar results were obtained by \citet{bzowski_etal_2012} (see Fig.~9 from their paper).
Note that our model results with the ``old'' and ``new'' LISM parameters do not precisely coincide with the results of \citet{bzowski_etal_2012} (see their Fig.~9) for several possible reasons. \citet{bzowski_etal_2012} included a
more detailed consideration of the geometry of IBEX measurements (e.g., taking into account the position and velocity of the spacecraft relative to the Earth, and a detailed consideration of the collimator shape and transmission function) and used more sophisticated helium ionization rates. Also, \citet{bzowski_etal_2012} applied an averaging of fluxes over each IBEX orbit, while
we performed calculations for all days during the orbits without averaging in time. Finally, the LISM parameters obtained in \citet{bzowski_etal_2012}
as the best fit of IBEX data are slightly different from our ``new'' LISM parameters, because we take these parameters from \citet{mccomas_etal_2012},
who quote compromise values based on the results of \citet{bzowski_etal_2012} and \citet{moebius_etal_2012}. However, our results are very close to
those of \citet{bzowski_etal_2012}, and differences between them are much smaller than differences between models
with the ``old'' and ``new'' LISM parameters (especially for the peak width, on which we are focused). A more detailed comparison of our results with
the model of \citet{bzowski_etal_2012} is outside the scope of this paper.

In order to increase the peak width for the model with the ``old'' LISM velocity vector we performed the calculations with an enhanced LISM temperature $T_{LISM}$=9000~K
(i.e. in the ``old with enhanced temperature'' model).
As seen from Fig.~\ref{gauss_ot_DOY}, this model leads to good agreement of peak widths with
the ``new model'' results. This means that, in principle, it is possible to fit the core of the helium fluxes measured by IBEX by increasing the LISM temperature instead
of changing the relative SW/LISM velocity vector. The association between a substantially higher LISM temperature with a smaller longitude ($\lambda_{LISM}$) and/or a higher magnitude of velocity ($|\textbf{V}_{LISM}|$) along the ``narrow tube'' in 4D space of the LISM
  parameters was also discussed in the original IBEX analyses \citep{moebius_etal_2012,bzowski_etal_2012,mccomas_etal_2012}. (See Fig. S3 and S1 in the supplementary materials of \citet{mccomas_etal_2012}.) Thus, our results are consistent with previous conclusions of the IBEX-Lo team.

In the same way as we did in the previous section, we investigate how the results presented in this section depend on the ionization rates.
In order to do that we repeat the calculations for the model with the ``old'' LISM velocity vector, but for different ionization rates.
We performed the calculation for the time and latitudinally dependent photoionization rate as described in previous section and in Appendix B.
In addition we consider ``extreme'' cases of negligible ($\beta_{ph,E}=0$~s$^{-1}$) and very high ($\beta_{ph,E}=5\cdot 10^{-7}$~s$^{-1}$) ionization rates.

\begin{figure}
\epsscale{.80}
\plotone{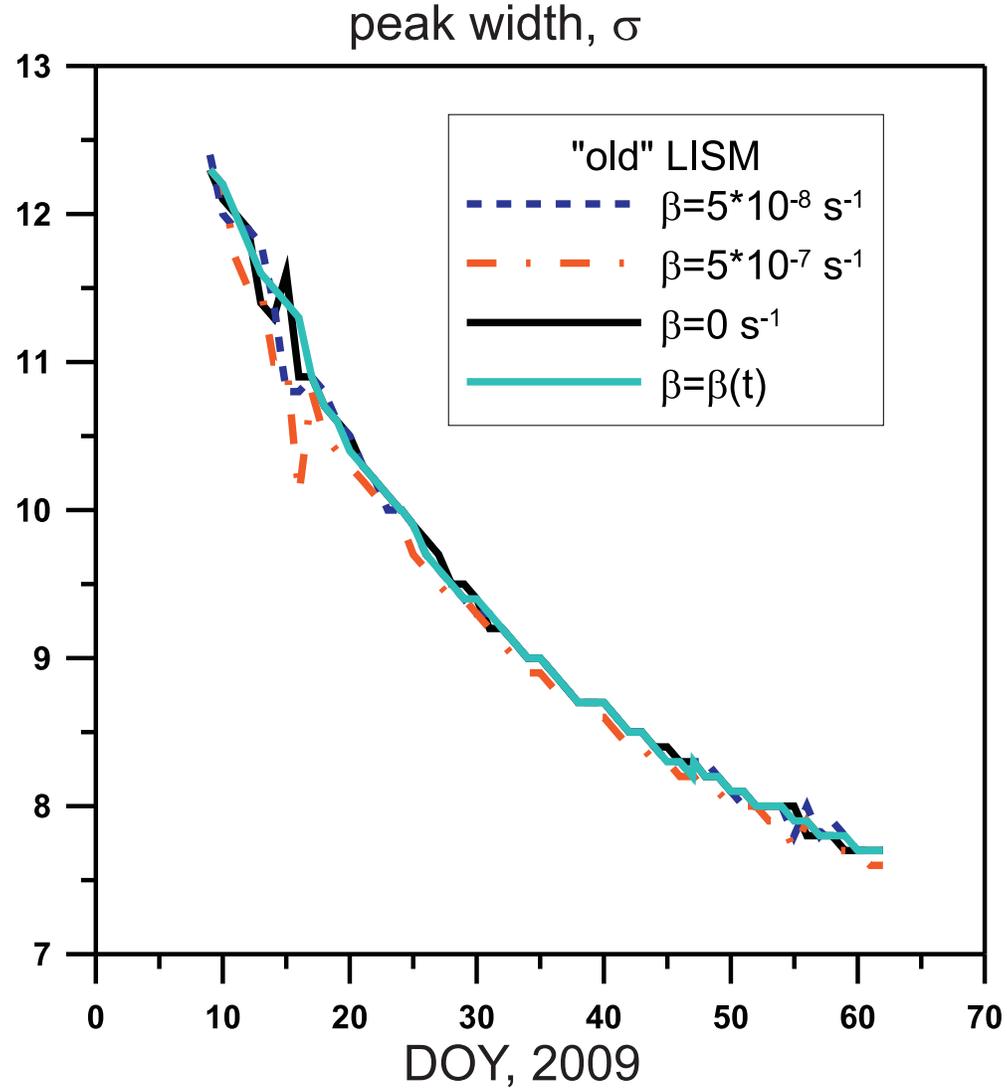}
\caption{ Peak width $\sigma$ as a function of DOY, obtained from calculations with the ``old'' LISM velocity vector for different photoionization rates.  The
solid black curve corresponds to the results with $\beta_{ph}=0$~s$^{-1}$, the dashed blue (black in printed version) curve corresponds to $\beta_{ph}=5\cdot 10^{-8}$~s$^{-1}$, the
dashed-dot orange (grey in printed version) curve corresponds to $\beta_{ph}=5\cdot 10^{-7}$~s$^{-1}$, and the solid light blue (light grey in printed version) curve corresponds to the results of a non-stationary model
with a time-dependent photoionization rate.
}\label{ibex_sigma_dif_betta}
\end{figure}

Fig.~\ref{ibex_sigma_dif_betta} shows the results for the peak width, and clearly demonstrates that this value does not depend on the ionization rate.
Similar results were obtained for the NEP-angle of the peak. We conclude that uncertainties in the photoionization rates are not a factor in the analysis.

\section{Conclusions and Discussion}

In this paper we have modeled the fluxes of the interstellar helium atoms measured by the Ulysses and IBEX spacecraft.
The calculations were performed for two different LISM velocity vectors: the ``old'' one from Ulysses/GAS 1990-2002 data \citet{witte_2004}
and the ``new'' one from IBEX-Lo 2009-2010 data \citet{mccomas_etal_2012}.

Our model results were compared with Ulysses/GAS maps of helium fluxes obtained in 2001 and 2007. The 2007 Ulysses/GAS data had not been analyzed before.
The comparison shows that the simulated theoretical maps agree with the Ulysses data fairly well for the models with the ``old'' LISM velocity vector.
At the same time, the model with the ``new'' LISM velocity vector cannot reproduce the correct direction of the interstellar helium flux maximum measured by Ulysses/GAS.

Simulations of the IBEX-Lo data with the ``old'' and ``new'' LISM velocity vectors have shown that the main difference in the model fluxes
lies in the ``peak width'', while the direction
of maximum fluxes (the NEP-angle $\alpha_{NEP,max}$) is in very good agreement between the two models. It is possible to change the peak width in the model by increasing the LISM temperature. These results are consistent with the original IBEX analyses \citep{moebius_etal_2012, bzowski_etal_2012, mccomas_etal_2012}
in terms of the ``narrow tube'' of LISM parameters in 4D space.


Therefore, we conclude that
\begin{enumerate}
\item Analysis of the Ulysses/GAS 2007 data shows that there is no change in the LISM velocity vector from that measured from the previously analyzed 1990-2002 Ulysses/GAS data.
\item It is impossible to get any reasonable agreement between model results and the Ulysses/GAS data
for the model with the ``new'' LISM velocity vector (i.e. the velocity vector derived from the IBEX-Lo data).
\item Contrary to Ulysses/GAS, for the IBEX-Lo observational geometry the directions of the flux maximum calculated in the frame of models with ``new'' and ``old'' LISM velocity vectors coincide fairly well.
The main difference between models with the two vectors lies in the width of the core of the helium signal.
\item These results do not depend on time and heliolatitudinal variations of the photoionization rate adopted in the models. Directions of maximum fluxes are also independent on the LISM temperature.
Thus, the results are very robust.
\item The width of the helium signal measured by IBEX depends on the LISM temperature and velocity, and an increase of the LISM temperature allows one to obtain
a good agreement in the  peak width between the model results with the ``old'' LISM velocity vector and the IBEX-Lo data.
\end{enumerate}

 Our analysis confirms that the LISM velocity vector derived from the IBEX data does not explain the Ulysses data and vice versa.
 We have shown that this conclusion does not depend on details (i.e. temporal and latitudinal variations) of the ionization rate. This means that the differences between LISM parameters derived by \citet{witte_2004} and \citet{bzowski_etal_2012} cannot be explained
  by differences in ionization rates adopted in their models.

We have shown that it is not possible to explain the Ulysses/GAS data with the ``new'' LISM velocity vector from IBEX by any means, because the direction of the He beam in the GAS data depends
only on $\textbf{V}_{LISM}$. At the same time, in principle, the IBEX data could be explained with the LISM velocity vector derived from the GAS data.
This is possible for models with the interstellar temperature increased from 6300~K to 9000~K.
However, the LISM temperature of 9000~K seems too high, since it strongly contradicts the width of Ulysses/GAS data as seen from the analysis of \citet{witte_2004}.
In addition, studies of local interstellar absorption features suggest a cooler interstellar temperature in the vicinity of the Sun \citep{redfield_etal_2008}.

Let us put the above discussion in the context of previous studies.
The velocity and temperature of the LISM have also previously been determined from the diffuse HeI-58.4 nm background radiation by Extreme Ultraviolet Explorer (EUVE)
and SOHO/UVCS \citep[see,][]{lallement_etal_2004, vallerga_etal_2004}.
 \citet{lallement_etal_2004} obtained the following LISM temperature and velocity vector in J2000 ecliptic coordinates: $T_{LISM}$=6500$\pm$2000,
$V_{LISM}=24.5 \pm 2$~km/s, $\lambda_{LISM}=75.4^{\circ} \pm 0.5^{\circ}$, $\beta_{LISM}=-5^{\circ} \pm 1^{\circ}$. The possible range of LISM temperature is less
than 9000~K, but the error bars are large. The same conclusion can be made for the absolute value of the LISM velocity. Error bars are large and include both Ulysses and IBEX derived values.
 However, the ecliptic longitude of vector $\textbf{V}_{LISM}$  obtained from the UV data analysis coincides very well with the value derived by \citet{witte_2004}.
 The given error band for $\lambda_{LISM}$ is quite small and the IBEX derived value is outside of it.

Besides the UV data, the ecliptic (J2000) longitude of the interstellar wind flow direction ($\lambda_{LISM}=75.13^{\circ} \pm 0.33^{\circ} $) was found previously from measurements of the interstellar helium pickup ions (PUIs) in 1998-2002 by ACE/SWICS \citep{gloeckler_etal_2004}.
This longitude of $\textbf{V}_{LISM}$ is very close to the results of \citet{witte_2004}, but it is 4$^{\circ}$ away from the new value
of IBEX. More recently, new measurements of PUIs performed in 2007-2011 by the PLASTIC instrument onboard the STEREO A spacecraft were used for determination
 of $\lambda_{LISM}$ by \citet{drews_etal_2012}. They used several techniques for obtaining the longitude of $\textbf{V}_{LISM}$ based on
 analysis of focusing ``cones'' \citep[similar to][]{gloeckler_etal_2004} and ``crescents'' in PUIs distribution, and in addition to
  He$^{+}$ ions they also analyzed Ne$^{+}$ and O$^{+}$. Final results of \citet{drews_etal_2012} with the least error bar correspond to the
  He$^{+}$ ``cones'' and give
 $\lambda_{LISM}=77.4^{\circ} \pm 1.9^{\circ} $, which is slightly closer to the ``new'' IBEX He vector than to the ``old'' Ulysses one.
 However, it should be noted that PUI distributions in the heliosphere are strongly affected by several processes, which may influence the position of the PUI focusing cone. One important process is the ``transport'' effect in the motion of PUIs due to anisotropies in the PUI velocity distribution \citep{moebius_etal_1995}, which can lead to angular displacement between the direction of the neutral helium flux and the helium PUI cone. The transport effect is more pronounced for time periods with low level of turbulence, which leads to larger mean free path of PUIs and, hence, more anisotropy in the PUI velocity distribution.
  Neither \citet{gloeckler_etal_2004}, nor \citet{drews_etal_2012} takes this effect into account in their analysis.
 However, \citet{chalov_fahr_2006} have shown that the ``transport'' effect does not greatly affect the results of \citet{gloeckler_etal_2004} due to particulars of the observations (namely the fact that the SWICS instrument detects only a fraction of the ions with certain velocity directions and magnitudes) and high level of turbulence during the solar maximum conditions. Measurements of helium pickup ions with STEREO were carried out during unusually quiet solar minimum (2007-2011 years) conditions, when the mean free path of pickup ions and, hence, anisotropy of their velocity distribution could be very large. It means that the transport effect may be important for these data and at least some estimations of this effect are necessary for correct interpretation of the results presented by \citet{drews_etal_2012}. Another process that may influence the position of the helium focusing cone is temporal short-scale modulations of the solar wind parameters. \citet{drews_etal_2012} show that this effect may lead to significant systematic bias of the focusing cone's axis. Temporal variations of the solar wind were not considered by \citet{gloeckler_etal_2004}, but \citet{drews_etal_2012} performed a specific analysis to take this effect into account under several assumptions. Thus, it is challenging to use measurements of PUIs for precise determination of the direction of the interstellar wind, and one should be careful with interpretation of results based on analysis of PUI distributions.


\citet{bzowski_etal_2012} mentioned that the recent analysis of the LISM structure performed by \citet{redfield_etal_2008}
 (based on high spectral resolution observations of interstellar absorption lines in the UV observed by Hubble Space Telescope, and from the Ca~II
optical transition observed from the ground)
showed that the flow vector of the Local Interstellar Cloud (LIC) in good agreement with the results of IBEX measurement.
However, the IBEX data measure the very local interstellar medium parameters (say at 1000 AU from the Sun),
 while the analysis of \citet{redfield_etal_2008}
deals with an integration of LIC parameters along lengthy lines of sight.
Thus, there could in principle be a real discrepancy between the truly local vector and the average vector measured towards nearby stars.

\citet{frisch_etal_2013} discussed short time scale (order of 10 years) variations in interstellar parameters as a
possible reason for the differences between the LISM velocity vector derived from the Ulysses/GAS and IBEX-Lo data.
Based on a linear fit of all previously published results about the direction of the LISM velocity vector and their uncertainties
\citet{frisch_etal_2013} stated that an increase of $\lambda_{LISM}$ over 40 years
is more likely than a constant flow direction.
A detailed discussion of many of the data used by \citet{frisch_etal_2013} is presented by \citet{lallement_bertaux_2014}, who end up questioning the conclusions of \citet{frisch_etal_2013}.
The analysis of Ulysses/GAS 1990-2002 data by \citet{witte_2004} and our analysis here of the 2007 data do not show any evidence for variation in the LISM vector within the 17 year lifespan of Ulysses.

The question why the LISM parameters derived from IBEX-Lo observations disagree with those derived from Ulysses/GAS data remains open.
If there are no systematic instrumental uncertainties in both Ulysses and IBEX data, then, in our opinion the only option is to look into new physical processes not considered
before in the models that could modify (increase) the width of the Maxwellian core of interstellar helium signal measured by IBEX at 1~AU. Simultaneously this (unknown)
effect should not influence the helium fluxes at larger (2-5~AU) distances as they are measured by Ulysses.


\acknowledgments

Calculations of helium fluxes and comparison with IBEX data was supported by RFBR grant No.~14-02-00746.
O.K. and V.I. would like to thank ISSI for their support of the working team 223.
B.W. acknowledges support from NASA award NNH13AV19I to the Naval Research Laboratory.
The reported study was performed by means of the Supercomputing Center of Lomonosov Moscow State University \citep{voevodin_etal_2012}.

\appendix

\section{Details of simulation of the helium fluxes measured by Ulysses/GAS}

In this section the procedure for calculating helium fluxes measured by GAS is discussed.
This section is largely based on \citet{banaszkiewicz_etal_1996}, where the
simulation of GAS measurements is described in detail.

The GAS instrument has two similar detectors with different angular resolution. The half cone angle $\theta_{max}$ is $\pm 3.7^{\circ}$
 for the wide field of view (WFOV) detector and $\pm 2.45^{\circ}$ for the narrow field of view (NFOV) detector. The effective area $S$ of the detector for
 particles entering along the optical axis (at $\theta=0^{\circ}$)
is $S_0=0.0908$~cm$^2$, while for other directions $S=S_0 \cdot G(\theta)$, with the geometric instrument function $0\leq G(\theta) \leq 1$.
Plots of the functions $G(\theta)$ for NFOV and WFOV are presented in Fig.~3 of \citet{banaszkiewicz_etal_1996}.

The probability of a particle's detection depends on the energy relative to the spacecraft ($E_{rel}$).
This probability is described by an
energy dependent efficiency function $f_{eff}(E_{rel})$.  A plot of this function is presented in Fig.~1 of \citet{banaszkiewicz_etal_1996}.

The spin axis of the Ulysses spacecraft is always oriented towards Earth. There is a spacecraft coordinate system defined by the positions of the spacecraft, Earth, and Sun \citep[see Fig.~2 in][]{banaszkiewicz_etal_1996}.
In this coordinate system there are two spherical angles, elevation ($\epsilon$) and azimuth ($\alpha$), which determine the direction of the line of
sight. In each GAS data file the transformation matrix from the spacecraft coordinate system to the solar ecliptic coordinate system is provided.
We use that to transform the ecliptic longitude and latitude to the elevation and azimuth.

During one rotation period the instrument scans a ring (or part of the ring) on the celestial sphere at a given elevation angle. The step of
 azimuth angle $\Delta \alpha$ is different for different scanning regimes and is equal to one of the following values: 0.7$^{\circ}$,
 1.4$^{\circ}$, 2.8$^{\circ}$, 11.2$^{\circ}$. Every 68 minutes the elevation angle changes by step $\Delta \epsilon$, which can be equal
 to 1$^{\circ}$, 2$^{\circ}$, 4$^{\circ}$ or 8$^{\circ}$. Thus, the scanned field is divided by a number of cells with angular resolution
 $\Delta \epsilon \times \Delta \alpha$. Counts measured per 100 seconds for each cell are provided in each GAS data file.

In order to simulate measured counts ($C_{i,j}$) in the chosen direction of the line of sight (defined by angles $\epsilon_i$ and $\alpha_j$) we need to take into account all effects mentioned above (field of view of the collimator, function $G(\theta)$, and energy efficiency $f_{eff}(E_{rel})$). Namely:
\begin{equation}
  C_{i,j} = S_0 \int_{\Omega(\epsilon_i, \alpha_j)} G(\theta) f_{eff}(E_{rel})
  f_{He}(\textbf{w}) |w_{rel}| d \textbf{w}_{rel}. \label{counts_1}
\end{equation}
Here, $f_{He}$ is the velocity distribution function of helium atoms, $\textbf{w}_{rel}=\textbf{w}-\textbf{V}_{SC}$ is the velocity of the atom relative to the spacecraft, $E_{rel}=m_{He}w_{rel}^2/2$ is the energy of each atom relative to the spacecraft, and
$\Omega(\epsilon_i, \alpha_j)$ is an integration area in velocity space related to the collimator. Specifically, if an atom's velocity vector $\textbf{w}_{rel}$ belongs to $\Omega(\epsilon_i, \alpha_j)$,
 then it is detected.

 In order to define the integration area let us consider a spherical coordinate system in velocity space.  In other words, let us describe velocity vector $\textbf{w}_{rel}$ by its magnitude $|w_{rel}|$ and two angles $\theta$ and $\phi$. The $\theta$ angle is measured
 from the optical axis of the collimator ($\theta=0^{\circ}$ corresponds to the center line in field of view) and $\theta \in [0, \theta_{max}]$,
 $\phi \in [0, 2 \pi]$. Thus,
 \[
 d \textbf{w}_{rel} = |w_{rel}|^2 sin(\theta) d |w_{rel}| d\theta d\phi,
\]
and
\begin{equation}
 C_{i,j} = S_0 \int_0^{+\infty} \int_0^{\theta_{max}}
 \int_0^{2\pi} G(\theta) f_{eff}(E_{rel})
  f_{He}(\textbf{r},\textbf{w}) |w_{rel}|^3 sin(\theta) d |w_{rel}| d\theta d\phi.
\end{equation}\label{eq_counts}
The dimensions of $C_{i,j}$ is $s^{-1}$, so this is in counts measured per second.

Knowing the velocity distribution function of helium at the location of Ulysses from solution of the kinetic equation~(\ref{Boltzmann}), we model the theoretical counts and compare them with the GAS data.

We chose two GAS maps for the simulation and comparison, one map in 2001 (day of year -- DOY=250) and one map in 2007 (DOY=251).
As extensively discussed in \citet{banaszkiewicz_etal_1996}, there is a background in the GAS data due to
contamination by EUV photons, cosmic rays, heavy elements and so on. We inspected other GAS maps in addition to the chosen two, and we did not see any evidence for any strong, localized background that would be in the He beam location in the chosen maps. Background concerns are a good reason to consider maps
from different parts of the sky in the analysis, and the two chosen maps are indeed in different parts of the sky due to the different Ulysses orbital motion.
\citet{banaszkiewicz_etal_1996} mentioned that the background can anisotropic due to contribution of heavier interstellar elements like oxygen and neon, because their spatial distribution is not uniform. In order to evaluate the influence of heavier elements, we performed the calculations of oxygen (O) and neon (N) fluxes in the frame of our numerical model for one chosen moment of time. We used the number densities of these species in the LISM from
\citet{izmod_etal_2004} and the ionization rates at 1~AU from \citet{cummings_etal_2002}. In these calculations we used the same energy efficiency function as for helium atoms \citep[presented by Fig. 1 in][]{banaszkiewicz_etal_1996}, as we have no other information on GAS
  efficiency for other elements. We found that fluxes of interstellar oxygen and neon are several orders smaller than interstellar helium fluxes. Thus, we can neglect them and assume that the background is spatially uniform,
i.e. it is just a constant for each map. The following values of the background were measured from the GAS data and added artificially to all results
of the modelling: $C_{background}=16.97$~counts/100~sec in 2001 (DOY=250) and $C_{background}=19.36$~counts/100~sec in 2007 (DOY=251).



\section{Three-dimensional time-dependent photoionization rate}

In this section the three-dimensional (3D) time-dependent treatment of the photoionization rate is described.
\citet{auchere_etal_2005a, auchere_etal_2005b} showed that the latitudinal distribution of the extreme-ultraviolet (EUV) solar flux is largely anisotropic.
This is due to the nonuniform distribution of bright features (active regions) and dark features (coronal holes) on the surface of the Sun. \citet{auchere_etal_2005a, auchere_etal_2005b} have developed a three-dimensional model for the He~II 30.4 nm flux observed at any heliospheric position
from January 1996 to August 2003. Their results were based on daily SOHO EUV Imaging Telescope (EIT) images. A detailed description of the method can be found
in \citet{auchere_etal_2005a}.

\begin{figure}
\epsscale{1.1}
\plotone{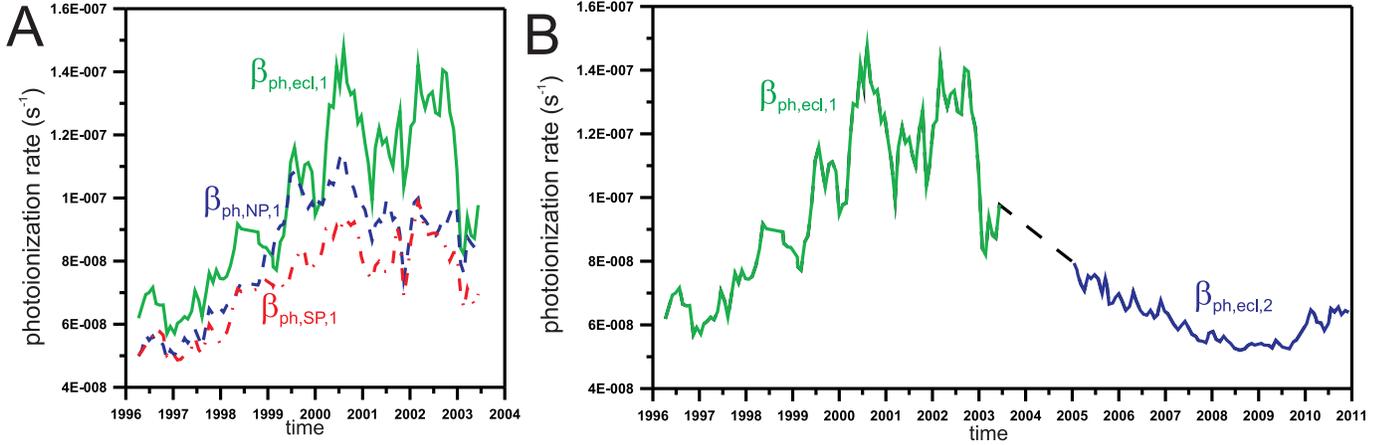}
\caption{A: The photoionization rate at 1~AU in the ecliptic plane ($\beta_{ph,ecl,1}$, solid curve), at the north pole ($\beta_{ph,NP,1}$, dashed curve) and at the
south pole ($\beta_{ph,SP,1}$, dashed-dot curve) as function of time obtained from SOHO/EIT data \citep{auchere_etal_2005a, auchere_etal_2005b}.
B: The photoionization rate in the ecliptic plane as a function of time for the full time period from 1996 to 2011.
The first part ($\beta_{ph,ecl,1}$, green or grey line) before the middle of 2003 corresponds to the results of \citet{auchere_etal_2005a, auchere_etal_2005b};
the last part ($\beta_{ph,ecl,2}$, blue or black line) from 2005 to 2011 corresponds to the results of \citet{bzowski_etal_2012};
 the middle part between them (dashed black curve) is just a straight line, because we assume a linear interpolation
 of the photoionization rate for this period of time. }\label{bph}
\end{figure}

\citet{floyd_etal_2012} calculated the photoionization rates of helium at 1~AU for the solar ecliptic plane ($\lambda\approx 0^{\circ}$) as well as for the north ($\lambda=+90^{\circ}$) and south ($\lambda=-90^{\circ}$) heliographic poles using
the results of \citet{auchere_etal_2005a, auchere_etal_2005b}.
In our work we used these photoionization rates from the beginning of 1996 to August, 2003 (see Fig.~\ref{bph}).
For heliolatitudes between $0^{\circ}$ and $\pm 90^{\circ}$ we assume a simple linear interpolation. Unfortunately, there are no available results of the Auchere model after 2003.


 The photoionization rate at 1~AU in the ecliptic plane from 2005 to 2011 is presented (and plotted) in \citet{bzowski_etal_2012}.
 It was obtained from the integration of the solar spectrum measured by TIMED/SEE \citep{woods_etal_2005} with the photoionization cross section from \citet{verner_etal_1996}.
Fig.~4 from \citet{bzowski_etal_2012} presents the time series (2005-2011) of Carrington period-averages of the photoionization rate of neutral helium at a distance of 1 AU from the Sun. We digitized this plot and obtained from it photoionization rates after 2005.
Between the middle of 2003 and the beginning of 2005 we assume a linear interpolation in time for the photoionization rate (see Fig.~\ref{bph}~B).
We do not assume any heliolatitudinal anisotropy of the photoionization rate after August 2003, because we have no appropriate data for it.
This means that during the period from 2003 to 2009 the model is time-dependent, but axisymmetric.
However, to check the influence of heliolatitudinal anisotropy of the ionization rate we performed specific test calculations. Namely, we calculated helium fluxes with a 3D stationary model with artificially included very strong heliolatitudinal anisotropy of the ionization rate.
We found that positions of the He beam are not changed in these test results. This confirms that heliolatitudinal anisotropy
is not a critical issue for our study.

\section{Details of the modelling of the IBEX-Lo data}

Here details of the simulation of the IBEX-Lo data are presented.
In our calculations of the IBEX-Lo data it was assumed that the position and velocity of IBEX coincide with those of the Earth. Also we neglect the elliptical shape of the Earth's orbit
around the Sun, and assume that the Earth's velocity vector is perpendicular to the Earth-Sun line of sight. The magnitude of the Earth's velocity
is approximated as constant and equal to 29.78~km/s.
Note that \citet{bzowski_etal_2012} have mentioned that ellipticity of Earth's orbit leads to the small radial component of the
Earth's velocity (on the order of 1 km/s ). And also of the order of a few km/s is the proper
motion of IBEX relative to the Earth. However, \citet{bzowski_etal_2012} studied this effect and have shown (see Fig.~13 from their paper) that the proper motion of the spacecraft around the Earth has the strongest influence (about 0.1~\%) only on the direction (NEP angle or spin angle) of the observed helium beam. For the peak height and especially for the peak-width, which we are particulary interested in here, the effect is negligible.
This means that our simplifications of the IBEX's and Earth's trajectories are appropriate for the purposes of this work.
Directions of the IBEX spin-axis for all dates of simulations are taken from the ISOC database \citep{schwadron_etal_2009}.

\begin{figure}
\epsscale{.80}
\plotone{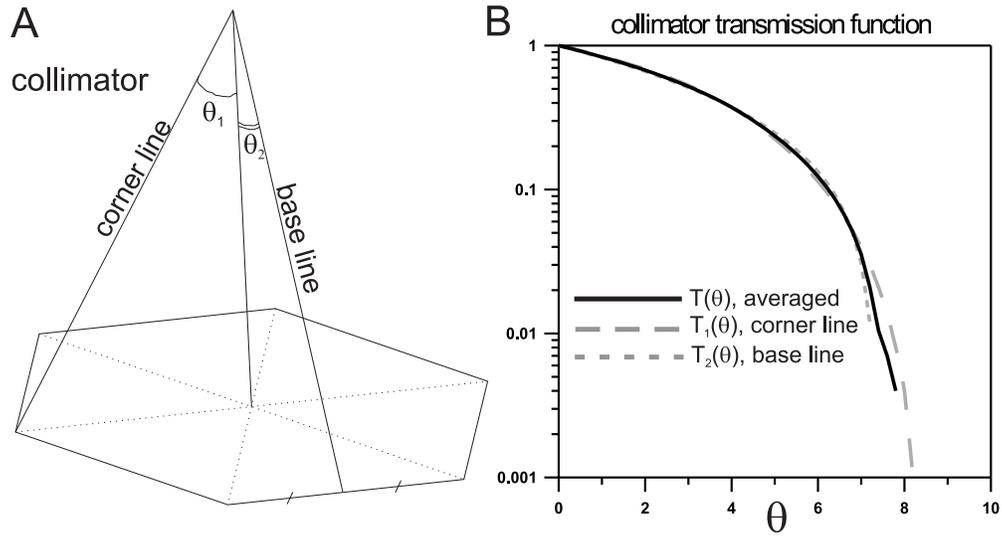}
\caption{ A: A hexagonal shape of the collimator used by \citet{bzowski_etal_2012}. Different transmission functions were adopted for the corner and
base lines of the hexagon; $\theta_1=8.4^{\circ}$, $\theta_2=7.4^{\circ}$. In our calculations we use the collimator with the conic form with the
  averaged cone angle $\theta_{max}=(\theta_1+\theta_2)/2=7.9^{\circ}$.  B: Collimator transmission functions $T_1(\theta)$ for the corner line and $T_2(\theta)$ for the base line (taken from
the ISOC database) and averaged transmission function $T(\theta)$ using in this work are presented.
}\label{ibex_collimator}
\end{figure}

\citet{bzowski_etal_2012} used the collimator with hexagonal field of view (FoV) and different collimator transmission functions at the corner and the baseline of hexagon \citep[see Fig.~2 from][]{bzowski_etal_2012}. The collimator transmission function of IBEX ($T(\theta)$) is analogous to the geometric instrument function $G(\theta)$ of GAS.
 The transmission function describes the probability of detection of the particles, which go through the collimator at an angle $\theta$ off the boresight axis.
 In our calculations we use for simplicity a circular FoV with an averaged angular resolution of $\theta_{max}=7.9^{\circ}$. And the collimator transmission function
 is taken as an average between transmissions at the corner ($T_{1}(\theta)$) and base line ($T_{2}(\theta)$) of the original hexagon (see Fig.~\ref{ibex_collimator}). These transmission functions ($T_1$ and $T_2$) were taken from the ISOC database.


 Fluxes of the interstellar helium atoms averaged over IBEX's collimator were calculated in the context of our kinetic model. The formula for the fluxes in a
  chosen direction is similar to the one used for the counts measured by Ulysses/GAS, namely:
\begin{equation}
 F_{coll} = \int_0^{+\infty} \int_0^{\theta_{max}}
 \int_0^{2\pi} T(\theta)f_{He}(\textbf{r},\textbf{w}) |w_{rel}|^3 sin(\theta) d |w_{rel}| d\theta d\phi.
\end{equation}\label{eq_fluxes}
Here, the integration over the velocity in the spacecraft reference frame ($\textbf{w}_{rel}=\textbf{w}-\textbf{V}_{Earth}$) is performed from zero to infinity without any limitations and energy response functions. \citet{moebius_etal_2012} describe how IBEX does not measure the incoming helium atoms directly,
but it measures the sputtered negative ions (H, C and O) in all energy bands below the energy of the incoming neutral atoms. It means that the original energy of the neutral He cannot be determined from the IBEX measurements. There is no information on the energy response function for the IBEX-Lo sensor, so this is not considered.

\clearpage



\clearpage

\begin{table}[!h]
 \vspace{6mm}
 \centering
  \caption[]{Comparison between the GAS data and the model results in axisymmetric stationary case: direction of the center of the He beam}\label{tab1}
\vspace{5mm}\begin{tabular}{|l|c|c|c|c|c|c|c|c|}
\hline
 & \multicolumn{2}{c|}{GAS data} & \multicolumn{3}{c|}{model with ``old'' LISM} & \multicolumn{3}{c|}{model with ``new'' LISM} \\
\hline year/DOY & $\epsilon_0$ & $\alpha_0$ & $\epsilon_0$ & $\alpha_0$ & $\gamma$  & $\epsilon_0$  & $\alpha_0$  & $\gamma$ \\
\hline 2001/250 & 47.30 & 192.15 & 48.4 & 192.0 & 1.10 & 45.6 & 189.2 & 2.73 \\
\hline 2007/251 & 37.28 & 232.02 & 37.0 & 231.4 & 0.47 & 33.0 & 232.6 & 4.29  \\
\hline
\end{tabular}
\end{table}






\clearpage


\begin{thebibliography}{}
\bibitem[Auch\`{e}re et al.(2005a)]{auchere_etal_2005a} Auch\`{e}re, F. et al.  2005a, ApJ, 625, 1036.

\bibitem[Auch\`{e}re et al.(2005b)]{auchere_etal_2005b} Auch\`{e}re, F. et al.  2005b, in ESA SP-592: Solar Wind
11/SOHO 16, Connecting Sun and Heliosphere, 327.

\bibitem[Banaszkiewicz et al.(1996)]{banaszkiewicz_etal_1996} Banaszkiewicz, M., M., Witte, H., \& Rosenbauer, H.  1996, A\&A Suppl. Ser.,
120, 587.

\bibitem[Baranov \& Malama(1993)]{bm_1993} Baranov, V. B., \& Malama, Yu. G.  1993, J. Geophys. Res., 98, 15157.

\bibitem[Bertaux \& Blamont(1971)]{bertaux_blamont_1971} Bertaux, J. L., \& Blamont, J.  1971, A\&A, 11, 200.

\bibitem[Bertaux et al.(1985)]{bertaux_etal_1985} Bertaux, J. L. et al.  1985, A\&A, 150, 1.

\bibitem[Bzowski et al.(2012)]{bzowski_etal_2012} Bzowski, M. et al.  2012, ApJS, 198, 12.

\bibitem[Chassefiere et al.(1986)]{chassefiere_etal_1986} Chassefiere, E., Bertaux, J. L., \& Sidis, V.  1986, A\&A, 169, 298.


\bibitem[Chalov \& Fahr(2006)]{chalov_fahr_2006} Chalov, S. V., \& Fahr, H. J., 2006, Astron. Let., 32, 487.

\bibitem[Costa et al.(1999)]{costa_etal_1999} Costa, J. et al.  1999, A\&A, 349, 660.

\bibitem[Cummings et al.(2002)]{cummings_etal_2002} Cummings, A. C., Stone, E. C., \& Steenberg, C. D., ApJ, 578, 194.

\bibitem[Drews et al.(2012)]{drews_etal_2012} Drews, C. et al.,  2012, Geophys. Res. Let., 40, A09106.


\bibitem[Gloeckler et al.(2004)]{gloeckler_etal_2004} Gloeckler, G. et al.,  2004, A\&A, 426, 845.



\bibitem[Gurnett et al.(2013)]{gurnett_etal_2013} Gurnett, D. A. et al.,  2013, Science, 341, 1489.

\bibitem[Floyd et al.(2012)]{floyd_etal_2012} Floyd, L. E., McMullin, D. E., Auchere, F.,  2012, AGU Fall meeting 2012 (SH11B-2212).


 \bibitem[Frisch et al.(2013)]{frisch_etal_2013} Frisch, P. C., et al.,  2013, Science, 341, 1080.

\bibitem[Fuselier et al.(2009)]{fuselier_etal_2009} Fuselier, S. A. et al.,  2009, Space Sci. Rev., 146, 117.

\bibitem[Izmodenov et al.(2000)]{izmod_etal_2000} Izmodenov, V. V. et al.,  2000, Astrophys. Space Sci., 274, 71.

\bibitem[Izmodenov et al.(2000)]{izmod_etal_2001} Izmodenov, V. V., Gruntman, M., Malama, Yu.G.,  2000, J. Geophys. Res., 106, 10681.

 \bibitem[Izmodenov et al.(2004)]{izmod_etal_2004}  Izmodenov, V. V., et al.,  2004, A\&A, 414, L29.

 \bibitem[Judge et al.(1998)]{judge_etal_1998} Judge, D. L., et al.  1998, Sol. Phys., 177, 161.

 \bibitem[Katushkina \& Izmodenov(2010)]{kat_izmod_2010} Katushkina, O. A. \&  Izmodenov V. V.,  2010, PAZh, 36, 297.

\bibitem[Katushkina \& Izmodenov(2011)]{kat_izmod_2011} Katushkina, O. A. \& Izmodenov V. V.,  2011, Adv. in Space Res., 48, 1967.

\bibitem[Katushkina et al.(2014)]{katushkina_etal_2014} Katushkina, O. A., Provornikova, E. A. \& Izmodenov V. V.,  2014, Astron. Let., 40, 135.

\bibitem[Lallement et al.(2010)]{lallement_etal_2010} Lallement, R. et al.,  2010, AIP Conference Proceedings, 1216, 555.

\bibitem[Lallement \& Bertaux(2014)]{lallement_bertaux_2014} Lallement, R. \& Bertaux J.-L.,  2014, A\&A, in press (arXiv:1402.1977, astro-ph.GA).

\bibitem[Lallement et al.(1985)]{lallement_etal_1985}  Lallement, R. et al.,  2004, A\&A, 426, 875.

\bibitem[Lallement et al.(2004)]{lallement_etal_2004}  Lallement, R. et al.,  2004, A\&A, 426, 875.

\bibitem[Lallement et al.(2005)]{lallement_etal_2005}  Lallement, R. et al.,  2005, Science, 307, 1447.

\bibitem[Lee et al.(2012)]{lee_etal_2012}  Lee, M. A. et al.,  2012, ApJS, 198, 10.

\bibitem[McComas et al.(2009)]{mccomas_etal_2009} McComas, D. J. et al.,  2009, Science, 326, 959.

\bibitem[McComas et al.(2012)]{mccomas_etal_2012} McComas, D. J. et al.,  2012, Science, 336, 1291.

\bibitem[Meier (1977)]{meier_1977} Meier, R. R.,  1977, A\&A, 55, 211.

\bibitem[M\"{o}bius et al.(1995)]{moebius_etal_1995} M\"{o}bius, E. et al.,  1995, A\&A, 304, 505.

\bibitem[M\"{o}bius et al.(2004)]{moebius_etal_2004} M\"{o}bius, E. et al.,  2004, A\&A, 426, 897.

\bibitem[M\"{o}bius et al.(2009)]{moebius_etal_2009} M\"{o}bius, E. et al.,  2009, Science, 326, 969.

\bibitem[M\"{o}bius et al.(2012)]{moebius_etal_2012} M\"{o}bius, E. et al.,  2012, ApJS, 198, 11.

\bibitem[McMullin et al.(2004)]{mcmullin_etal_2004} McMullin, D. R. et al.,  2004, A\&A, 426, 885.

\bibitem[Quemeraiset al.(2010)]{quemerais_etal_2010} Quemerais, E. et al,  2010, ApJ, 711, 1257.

\bibitem[Quemerais \& Izmodenov(2002)]{quemerais_izmod_2002} Quemerais, E. \& Izmodenov, V. V.,  2002, A\&A, 396, 269.

\bibitem[Redfield \& Linsky(2008)]{redfield_etal_2008} Redfield, S. R., Linsky, J. L.,  2008, ApJ, 673, 283.

\bibitem[Saul et al.(2012)]{saul_etal_2012} Saul, L., et al.,  2012, ApJS, 198, 14.

\bibitem[Schwadron et al.(2009)]{schwadron_etal_2009} Schwadron, N. A., et al.,  2009, Space Sci. Rev., 146, 207.

\bibitem[Schwadron et al.(2013)]{schwadron_etal_2013} Schwadron, N. A., et al.,  2013, ApJ, 775, 86.

\bibitem[Thomas \& Krassa(1971)]{thomas_krassa_1971} Thomas, G. \& Krassa, R.,  1971, A\&A, 11, 218.

\bibitem[Vallerga et al.(2004)]{vallerga_etal_2004} Vallerga, J. et al.,  2004, A\&A, 426, 855.

\bibitem[Verner et al.(1996)]{verner_etal_1996} Verner, D. A., Ferland, G. J., Korista, T. K., Yakovlev, D. G.,  1996, ApJ, 465, 487.

\bibitem[Vincent et al.(2011)]{vincent_etal_2011} Vincent, F. E. et al,  2011, ApJ, 738, 10.

\bibitem[Voevodin et al.(2012)]{voevodin_etal_2012} Voevodin, Vl. V. et al,  2012, Open Systems J., 7, 36 (in russian).

\bibitem[Witte et al.(1993)]{witte_etal_1993} Witte, M., Banaszkiewicz, M., Rosenbauer, H.,  1993, Adv. Space Res., 13, 121.

\bibitem[Witte et al.(1996)]{witte_etal_1996} Witte, M., Banaszkiewicz, M., Rosenbauer, H.,  1996, Space Sci. Rev., 78, 289.

\bibitem[Witte(2004)]{witte_2004} Witte, M.,  2004, A\&A, 426, 835.

\bibitem[Woods et al.(2005)]{woods_etal_2005} Woods, T. N. et al.,  2005, J. Geophys. Res., 110, A01312.

\bibitem[Wu \& Judge(1979)]{wu_judge_1979} Wu, F. M. \& Judge, D. L.,  1979, ApJ, 231, 594.

\bibitem[Zank et al.(2013)]{zank_etal_2013} Zank, G. P. et al.,  2013, ApJ, 763, id 20.

\end{thebibliography}
\end{document}